\documentclass[reprint,eqsecnum,floats,aps,amsmath,amssymb,nofootinbib,prd,onecolumn, showpacs]{revtex4-1}

\usepackage{graphicx}
\usepackage{amsmath,amssymb}
\usepackage{hyperref}
\usepackage{graphicx}
\usepackage{subfigure}
\usepackage{arydshln}

\usepackage{graphicx}
\usepackage{amsmath,amssymb}
\usepackage{hyperref}

\begin{document}

\title{Perturbations in Quantum Cosmology: The Continuum Limit in Fourier Space}

\author{Beatriz Elizaga Navascu\'es}
\email{beatriz.b.elizaga@gravity.fau.de}
\affiliation{Institute for Quantum Gravity, Friedrich-Alexander University Erlangen-N{\"u}rnberg, Staudstra{\ss}e 7, 91058 Erlangen, Germany}
\author{Guillermo  A. Mena Marug\'an}
\email{mena@iem.cfmac.csic.es}
\affiliation{Instituto de Estructura de la Materia, IEM-CSIC, Serrano 121, 28006 Madrid, Spain}

\begin{abstract}
We analyze the passage to a continuum limit of the mode spectrum of primordial perturbations around flat cosmological spacetimes in Quantum Cosmology, showing that this limit can be reached even if one starts by considering a finite fiducial cell as spatial slice. Whereas the resulting system can be described in an invariant way under changes of the fiducial volume using appropriate variables, both for the background cosmology and the perturbations, obtaining in this way a discrete mode spectrum owing to the compactness of the fiducial cell, we show that the desired continuum limit for the perturbations can still be established by means of scaling transformations of the physical volume when this volume grows unboundedly. These transformations lead to a model with a continuum of modes and independent of any scale of reference for the physical volume. For the sake of comparison, we also consider an alternative road to the continuum in Fourier space that has been employed in geometrodynamics and is based on the use of scaling transformations of the fiducial volume, together with variables that are independent of them.  
\end{abstract}


\maketitle

\section{Introduction}

The standard cosmological model is based on the principle \cite{cosmoprinciple} that, over sufficiently large regions (of the order of 100 Mpc), our Universe is homogeneous in average. This principle is supported by observations of the large scale structure \cite{lss1,lss2,lss3} and is consistent with the isotropy of the cosmic microwave background (CMB) \cite{planck} (usually complemented with theoretical results about General Relativity (GR) for certain types of matter \cite{egs,egsapp}). This spacetime, homogeneous in average, contains inhomogeneous structures that can be explained by the evolution of primordial fluctuations in the early epochs of the Universe, amplified through a period of (exponentially) fast expansion that is known as inflation \cite{inflation,mukh}.  

In the attempt to construct a quantum formalism for cosmology, the coexistence of this homogeneity with inhomogeneous perturbations has posed some subtle tensions for models with noncompact spatial sections, because the integration over an entire spatial slice of the homogeneous degrees of freedom is proportional to the volume, that diverges in the noncompact case. Inhomogeneous perturbations, on the other hand, can be described as fields, that can be handled with due care over infinite sections of constant time. Notice that this kind of problems with integrals over spatial sections appear, in one way or another, in all approaches to cosmology constructed from an action, both with a Lagrangian or with a Hamiltonian formulation. Traditionally, this tension has been settled either by restricting all considerations to models with compact sections, for which the spatial volume is finite, or by absorbing the spatial volume in a redefinition of the action that rescales it. This latter line of action is in principle viable inasmuch as a global multiplicative factor in the action does not modify the dynamics or the constraint algebra. In the Hamiltonian formulation and for gravitational systems, that are totally constrained, it amounts to a scaling of the symplectic structure and of the Lagrange multipliers of the constraints (ignoring at this stage possible surface terms). For instance, for models in vacuo the spatial volume can be absorbed by a redefinition of the Newton constant, that divides the total action. This is commonly the procedure in cylindrically symmetric gravity or with plane symmetry, for the sake of some examples. In these cases, one adopts a redefined gravitational constant per unit length in the direction of the cylindrical axis, or per unit area of the symmetry plane, respectively \cite{cy1,cy2,cy3}. Out of the vacuum, the Newton constant provides the coupling with the matter content, and is not a multiplicative constant anymore. But the passage to the quantum theory carries with it the introduction of a global constant that makes the action dimensionless, e.g. to provide a phase that can be weighted in a path integral formulation, or simply to render dimensionless the symplectic structure (namely, the structure that determines the Poisson brackets). In most scenarios, this constant can be thought as the Planck constant. Absorbing the divergent spatial volume by redefining it might be considered a sort of renormalization.

However, one may argue that scaling the symplectic structure by a divergent quantity is not fully satisfactory. If one starts e.g.  with a canonical description of GR, one may judge problematic that its particularization to homogeneous and isotropic cosmological systems would lead to meaningful canonical relations only after a redefinition of a global constant that originally was assumed to be finite and universal. Rather, one might expect that the description of the system needs to be adapted to the homogeneous and isotropic reduction. To try and solve this objection along the sketched line of attack, a possible strategy is to cope with spatial integrals in a little bit more subtle way. To avoid divergences owing to infinite spatial volumes when considering homogeneous degrees of freedom, the integrals can be performed over a finite fiducial cell. For most practical purposes, this cell can be regarded as a compact spatial manifold. But then one must require that the physical results be independent of the choice of this cell.\footnote{Strictly speaking, it might suffice that the limit in which the cell becomes noncompact is well defined. However, it is more natural to demand complete invariance under the choice of cell, a property that guarantees the existence of the desired noncompact limit.}. This requirement can be achieved simply by redefining the basic variables (once specialized to the homogeneous and isotropic reduction) so that they are conveniently scaled by a power of the cell volume, in such a way that the resulting variables are genuinely canonical, with Poisson brackets that, in particular, are invariant under changes of the cell. For the sake of an example, this route has been followed in the application of Loop Quantum Gravity (LQG)\cite{lqg} to cosmology, a discipline that is called Loop Quantum Cosmology (LQC) \cite{ap,Edward}. LQG is a canonical program for the nonperturbative quantization of GR that uses as basic variables a densitized triad and a $su(2)$-connection. LQC has reached notable success in the study of homogeneous (and isotropic) universes \cite{aps,iaps,mmo}, among which probably the most renamed is the resolution of the Big Bang  singularity, which is replaced by a bounce, in which quantum geometry effects behave as a repulsive gravitational interaction \cite{ap,iaps}. In addition, primordial perturbations have also been considered in LQC (see e.g. the review in Ref. \cite{Edward}), and the possible consequences of a quantum geometry on the CMB have been studied \cite{dressed4,AMorris,AAG,hybrpred,hybrgui}. 

If,  according to this alternative strategy, one admits the restriction of the spatial sections to a fiducial cell, and imposes boundary conditions for the modes of the primordial perturbations that are consistent with this spatial restriction (e.g. periodic conditions), one gets a quantized collection of modes, with wavenumbers that form a discrete sequence. One would expect that the limit of a continuum of perturbative modes ought to be recovered in the limit of infinite volume for the fiducial cell. However, if the formulation is truly independent of this cell, as we have required, and this independence is maintained when the perturbations are included, it is obvious that one cannot attain this continuum by enlarging the fiducial cell unboundedly. One then may wonder how it is possible to define this continuum limit properly. The aim of this work is to clarify this issue and, moreover, show that the continuum limit in Fourier space is reached without the need to introduce any artificial scale in the system, something that would have spoiled the independence on background quantities that is characteristic of GR and seems desirable in any nonperturbative quantization of it, like LQG.

The modes of the gauge-invariant perturbations, namely the tensor modes and, e.g., the modes of the Mukhanov-Sasaki scalar (which are specially suitable in the case of flat spatial topology), propagate in the Early Universe subject to dynamical equations that can be written in GR like those of a harmonic oscillator with a time-dependent mass. In a large family of approaches to Canonical Quantum Cosmology \cite{HH,dressed1,dressed2,hybrinf3,hybrref,hybrmass}, the evolution equations of these modes ultimately remain of the same harmonic-oscillator type (at least in certain regimes), but the time-dependent masses change with respect to General Relativity, incorporating corrections that are due to quantum gravitational effects. These corrected masses are typically given by expectation values of some (few) operators on the part of the quantum state that describes the homogeneous geometry \cite{hybrref}. Furthermore, if this partial wavefunction displays a kind of effective behavior, in the sense that it remains peaked on a certain geometry during the whole of the evolution, instead of dealing with the quantum expectation values that determine the time-dependent masses of the perturbations, it should suffice to approximate them with their effective counterpart, equal to the evaluation on the peak trajectory. A frequent scenario is that the trajectory of this peak follows the Einsteinian dynamics except in certain regions with large quantum effects (see e.g. \cite{ap,kiefer}). Moreover, in cases like LQC, the new trajectory still follows a Hamiltonian dynamics, described by a Hamiltonian constraint that includes the quantum corrections in an effective way \cite{ap,AG}. 

The scale invariance of the equations of the perturbative modes for flat topologies in GR, when one goes to the Fourier continuum, suggests that the final result in Canonical Quantum Cosmology should be sensitive only to ratios of geometric scales in the proper continuum limit. Thus, one should be able to reach a limit with a continuum of modes such that the final physical equations don't display any dependence on the absolute value of the scale factor of the  {\sl background} homogeneous geometry, but rather on its ratio at two instants of time. A possible choice of reference is the usual one in cosmology: the present value of the scale factor. An equally valid possibility is the value of the scale factor at any definite instant of time, e.g. the onset of inflation or, in the case of LQC, when the bounce that replaces the Big Bang occurs. We notice that the existence of the two kind of scaling transformations that we are commenting, namely a scaling of the fiducial volume of the chosen cell, and a scaling of the scale factor of the geometry (which leaves invariant the ratio at two different times), were already pointed out in Ref. \cite{ap}, emphasizing the need to get a formulation with well-defined physical results in the limit in which the finiteness of the cell is removed.

The rest of this work is organized as follows. In Sec. 2 we present our cosmological system and its perturbations, summarizing well-known results both from GR and from its canonical quantization. We will provide the relevant formulas for the study of the background geometry and the gauge invariant perturbations. In addition, we will briefly explain the properties of GR that must be respected under quantization so that our analysis applies as well to Canonical Quantum Cosmology. In Sec. 3 we discuss the invariance of the quantum formulation under changes in the fiducial volume of the spatial sections, and investigate the behavior under the multiplication of the scale factor by a constant. In this way, we are able to prove that one can reach a well-defined continuum limit for the modes of the perturbations. Then, in Sec. IV we investigate an alternative route to this continuum, closer in spirit to a procedure adopted traditionally in geometrodynamics, namely we absorb the spatial volume in a redefinition of the fundamental action constant of the system (e.g. the Planck constant). We discuss some further aspects of our results and conclude in Sec. V. We set the speed of light equal to one, as well as the Planck constant or, eventually, its effective counterpart when properly explained.

\section{Mode equations}\label{secmode}

\subsection{The classical perturbed cosmological system}

Let us start by considering a flat, homogeneous, and isotropic model of the Fridemann-Lema\^{\i}tre-Robertson-Walker (FLRW) type, with spacetime metric of the form
\begin{equation}\label{flrwmetric}
ds^2 = - N_0^2(t) dt^2 +a^2(t) ~^0 h_{ij} d \theta^i d \theta^j ,
\end{equation}
where $N_0$ is the homogeneous lapse function, $a$ is the scale factor, and $~^0 h_{ij}$ is the Euclidean metric. Latin indices from the middle of the alphabet indicate spatial indices. In order to deal with spatial sections of finite fiducial volume, we assume the spatial topology of a three-torus, with angular spatial coordinates $\theta_i $ that have a period equal to a given constant $l_0$.\footnote{This corrects a misprint in Ref. \cite{hybrinf3} and related literature.} Note that the fiducial spatial volume is then ${\cal V}_0=l_0^3$. Thus, we will treat $l_0$ as a parameter of our description that we can change at will in order to modify the fiducial volume ${\cal V}_0$. The physical volume of the spatial sections, on the other hand, is just ${\cal V}_P=a^3{\cal V}_0$. Again, we notice that we can make the physical volume tend to infinity while keeping the fiducial one finite by letting the scale factor become unboundedly large. 

Notice that, in principle and for fixed Newton's constant, the symplectic structure and the Hamiltonian that generates the dynamics of the homogeneous system will scale linearly with the fiducial volume (essentially because both are defined so far as extensive quantities). Nevertheless, if one wants canonical relations and dynamics that do not vary with the choice of fiducial cell, one can introduce variables for the geometry that absorb the effect of the transformations in the fiducial element. A standard choice $({\cal V},B)$, that in fact differs from the definitions adopted in LQC only by multiplicative constants independent of the selected cell,\footnote{\label{Immi}The volume variable used in LQC is $v={\cal V}/(2\pi G \gamma \sqrt{\Delta_g})$, where $\gamma$ is the Immirzi parameter and $\Delta_g$ is the area gap, i.e. the minimum positive area eigenvalue that is allowed by LQG \cite{iaps}. On the other hand, the connection variable of LQC is $b= 4\pi G \gamma \sqrt{\Delta_g} B$.} is given by the physical volume up to a sign that determines the orientation of the triad, and by a variable that is proportional to the Hubble parameter: 
\begin{equation}
\label{ABvariables}
| {\cal V} |= a^3{\cal V}_0 = {\cal V}_P,\quad \quad {\cal V} B = - \frac{1}{3} a \pi_{a}.
\end{equation} 
We use the notation $\pi_x$ to denote a canonical momentum of the generic variable $x$, such that their Poisson bracket equals the unit. Besides, in order to provide a matter content to the model and render it nontrivial, we include a homogeneous scalar field, $\phi$, that will play the role of an inflaton, and will be subject to a potential $V(\phi)$. Then, the associated phase space can be described with two canonical pairs for any value of the fiducial volume: a pair for the FLRW geometry, $({\cal V},B)$, and another one for the inflaton, $( \phi , \pi_{\phi} )$. 

In absence of any matter content other than the homogeneous inflaton, the model is subject only to one constraint, which is a homogeneous, global Hamiltonian constraint $H_{|0}=0$, with Lagrange multiplier given by the lapse $N_0$, and
\begin{equation}\label{effh}
H_{|0}=\frac{1}{2 {\cal V} }\left[\pi_\phi^2- 12\pi G {\cal V}^2 B^2 + 2 {\cal V}^ 2 V(\phi) \right].
\end{equation}
In addition, the energy density $\rho$ and pressure $P$ of the inflaton are given by the sum and the difference, respectively, of the kinetic and the potential energy densities:
\begin{equation}\label{densitypressure}
\rho = \frac{\pi_{\phi}^2} {2{\cal V}^2 }+V(\phi) , \quad\quad P=\rho-2 V(\phi).
\end{equation}

Let us consider now small anisotropies and inhomogeneities, that we view as perturbations around our FLRW model, both in the geometry and the inflaton \cite{HH,kiefer,hybrinf2,hybrref,dressed2}. To handle the spatial dependence, we take advantage of the background symmetries and expand the perturbations in scalar, vector, and tensor harmonics of the spatial Laplacian, associated with the fiducial Euclidean metric \cite{hybrref}. Part of these perturbations are simply gauge degrees of freedom, that arise due to the possibility of modifying the original spacetime with a perturbative diffeomorphism. The physical information encoded in the perturbations resides in gauge-invariant quantities \cite{bardeen}. In the considered model, where the matter content is a scalar field, the only gauge invariants are the tensor modes of the perturbations and a linear combination of the scalar perturbations of the metric and the scalar field. In our case with flat spatial topology, a convenient choice of this scalar gauge invariant is the so-called Mukhanov-Sasaki (MS) scalar \cite{mukhanov2,sasaki,kodasasa}. 

In the description of these gauge invariant perturbations, one often uses the coefficients of their mode expansion in terms of harmonics: $T_{\vec{k}}^{(\epsilon)}$ and $Q_{\vec{k}}$, respectively for the tensor and the MS modes. Here, $\vec{k}$ is the wavevector of the mode, with its (Euclidean) norm $k$ providing the corresponding wavenumber. Besides, $\epsilon$ denotes the two possible polarizations of the gravitational tensor modes. A rescaling of these mode coefficients by the scale factor and by suitable powers of the fiducial volume leads to new mode variables \cite{hybrref} that indeed satisfy equations of the harmonic-oscillator type in GR:\footnote{These new coefficients differ by a factor of ${\cal V}_0^{-1/6}$ from those used in Refs. \cite{hybrgui,hybrmass}. In consonance, their respective canonical momenta are related with those in such references by a rescaling with ${\cal V}_0^{1/6}$.}
\begin{equation}
\label{ginvariantvariables}
{\bar d}_{\vec{k},\epsilon}= \frac{a T_{\vec{k}}^{(\epsilon)}}{ \sqrt{32\pi G} \, {\cal V}_0^{2/3}} \quad,\quad  {\bar v}_{\vec{k}}= \frac{a Q_{\vec{k}}}{ {\cal V}_{0}^{2/3}}.
\end{equation}
In the conformal time defined by the choice of lapse function $N_0=l_0 a= {\cal V}^{1/3}$, and with the corresponding time derivative denoted with a prime, one obtains the following mode equations according to the Einsteinian evolution:
\begin{eqnarray}\label{tensorGR}
{\bar d}_{ \vec{k},\epsilon}^{\prime\prime} + \left\{ {\cal V}_0^{2/3} k^2+   8 \pi G {\cal V}^{2/3} \left[ 2 \pi G B^2 -    V (\phi)\right]  \right\} {\bar d}_{\vec{k},\epsilon}&=& 0, \\
\label{MSGR}
{\bar v}_{ \vec{k}}^{\prime\prime}+ \left\{{\cal V}_0^{2/3} k^2+  8\pi G {\cal V}^{2/3} \left[ 2\pi G B^2 -   V(\phi) \right]  + U_{\rm MS} \right\} {\bar v}_{\vec{k}}&=& 0,
\end{eqnarray}
where the extra term in the equation of the scalar modes, that we will  call the MS potential, is given by the expression
\begin{equation}
\label{Mspotential}
U_{\rm MS}={\cal V}^{2/3}  \left[V_{,\phi \phi}(\phi)+ \frac{4 \pi_{\phi}} {{\cal V}B} V_{,\phi}(\phi)+48 \pi G V(\phi) - \frac{8  } {B^2} V^{2}(\phi)  \right]. 
\end{equation}
For the scalar field potential, the comma followed by $\phi$ denotes the derivative with respect to the inflaton. 
 
In the following we will use the name time-dependent mass to refer to the k-independent term that multiplies the mode variable in our equations of harmonic-oscillator type. The difference between the scalar and the tensor time-dependent masses is precisely the MS potential. Besides, a simple computation shows that the tensor mass can be rewritten in the alternative forms:
\begin{equation}
\label{tensormassderivatives}
8\pi G {\cal V}^{2/3} \left[ 2\pi G B^2 -  V (\phi)\right] = - \left\{ |{\cal V}|^{1/3} \left\{ |{\cal V}|^{1/3} ,H_{|0} \right\} ,H_{|0} \right\}= -\frac{a^{\prime\prime}} {a}.
\end{equation}
Here, curved brackets denote Poisson brackets.

At this point of our analysis, we emphasize that, if we insist in adopting the rescaled variables ${\cal V}$ and $B$, as well as the rescaled coefficients introduced for the perturbations, the only dependence on the fiducial volume that might survive in the mode equations is that appearing in the term that contains the square wavenumber. Nonetheless, this wavenumber is defined as the norm of the eigenvalues of the Laplace-Beltrami operator for the metric of the three-torus with fundamental period equal to $l_0={\cal V}_0^{1/3}$. Therefore, we get that $k=2 \pi |\vec{n}| {\cal V}_0^{-1/3}$ for any integer vector $\vec{n}$. As a consequence, the term ${\cal V}_0^{2/3}k^2$ is in fact also independent of the fiducial volume. However, we see that this term is quantized, namely, it only can equal the square norm of an integer vector multiplied by $4\pi^2$. Strictly speaking, if we then take the limit of infinite fiducial volume, the relevant mode spectrum remains discrete. Actually, this is what we should expect, because the formalism was constructed to be independent of the value of the fiducial volume, and for finite values the spectrum of the Laplace operator is not continuous. As we will show later in our discussion, this tension, that would put into question the passage to the continuum picture, can still be solved by adopting an alternative road to the Fourier continuum of perturbative modes. 

\subsection{Quantum corrections to the perturbed system}

In the framework of Canonical Quantum Cosmology \cite{HH,hybrref}, one adopts the following strategy to study our cosmological system. The action of the perturbed cosmology is truncated at quadratic perturbative order, regarding the homogeneous variables as zero modes that are treated exactly at that truncation order. The truncated system inherits a symplectic structure and constraints from GR \cite{HH}. In fact, one can find a canonical set of variables for the entire system so that the perturbations are described by the linear perturbative constraints, their associated gauge degrees of freedom, and a complete set of gauge invariants like, e.g., the tensor and the MS mode coefficients that we have been using in our discussion, together with canonical momenta for them \cite{langlois}. To preserve the canonical structure of the whole system, including the homogeneous sector, one needs to correct the zero modes with suitable quadratic contributions of the perturbations \cite{pintoneto1,pintoneto2,hybrref}. These corrections can be viewed as a kind of backreaction in our definitions. Replacing the original zero modes with the new ones, the treatment of the cosmological model and the formulation that we have explained in the previous subsection continues to be valid \cite{hybrref}. Furthermore, in those interesting situations where the backreaction is not relevant, the old and new zero modes may be identified without any physical consequence.  

Employing this canonical description, it becomes clear that physical quantum states can depend only on zero modes and perturbative gauge invariants. These states are still subject to a global constraint, that can be understood as the zero mode of the gravitational Hamiltonian constraint, truncated at quadratic perturbative order. It is given by the sum of a term that is formally identical to the homogeneous constraint $H_{|0}$ and the contributions of the tensor and MS Hamiltonians, that we will call $^{T}H_{|2}$ and $^{s}H_{|2}$, respectively, and that are quadratic in the gauge invariant perturbations \cite{hybrref}. We recall that these Hamiltonians generate the dynamical evolution of the tensor and MS perturbations in GR. In order to find solutions to this global constraint that can be of physical interest, it is convenient to introduce an ansatz of separation of variables, so that one can factorize the dependence of the quantum states on the FLRW geometry and on each of the different gauge invariant modes. In this ansatz, each partial wavefunction of the quantum state is allowed to depend on the inflaton, that is therefore employed as an internal time of the system to which one can refer the evolution of each part of the state. 

The part that contains the FLRW geometry can be quantized according to the specific rules that one adopts in Canonical Quantum Cosmology for the representation of the homogeneous gravitational sector. It is common to use a representation in which the volume variable ${\cal V}$ acts by multiplication. In particular, following these rules one promotes to operators the coefficients of the tensor and MS Hamiltonians that depend on the FLRW geometry. Besides, in the spirit that the gauge invariant modes are mere perturbations, it is usual to assume that the interaction with them does not produce significant changes in the quantum state of the FLRW geometry. With the ansatz of separation of variables and this assumption, it is not difficult to derive equations for the evolution of the gauge invariant perturbations \cite{hybrref}. The dynamical equations that one obtains for the perturbative modes in this way are very similar to those displayed in Eqs. \eqref{tensorGR} and \eqref{MSGR}, but with a modified time-dependent mass for the tensor and for the scalar gauge invariants. It is straightforward to convince oneself that, with a quantization strategy like the one that we have described, the new masses are in fact provided by (ratios of) expectation values of quantum operators on the quantum state of the homogeneous geometry. In this manner, quantum effects are incorporated into the dynamical description of the primordial perturbations. Moreover, from the fact that the classical time-dependent masses had been defined without any reference to the fiducial volume using homogeneous geometric variables (directly related to ${\cal V}$ and $B$), and as far as the quantization of these variables respects the independence on the fiducial structure (see e.g. \cite{hybrref}), it turns out that the situation with respect to this invariance is the same that we encountered above.

Let us focus on the scenario that seems more relevant for cosmological applications, namely, we will restrict our attention to quantum states of the FLRW geometry that are (approximate) solutions of homogeneous and isotropic Quantum Cosmology, so that the effect of the perturbations on them is negligible, i.e. the backreaction of the perturbations can be ignored. Furthermore, we will consider states that are peaked on effective trajectories, as we mentioned in the Introduction. Typically, these trajectories are generated, rather than by the classical constraint $H_{|0}$, by an effective counterpart of it, $ H_{|0}^{\rm eff}$, that one obtains by incorporating the effect of quantum gravitational corrections on the homogeneous sector.\footnote{\label{effect}An example is the effective Hamiltonian of LQC, obtained from that of GR by replacing $16 \pi^2 G^2\gamma^2 \Delta_g B^2$ with $\sin^2{(4\pi G \gamma \sqrt{\Delta_g} B)}$ \cite{ap,iaps}.}

For sufficiently concentrated states, a good approximation consists in substituting the quantum expectation values by their evaluation on the peak trajectory, i.e. on the effective solution. When one does so, the resulting time-dependent masses are similar to those of the classical case, except for the incorporation of quantum corrections owing to the following. First, now the homogeneous background does not necessarily follow a GR trajectory, but rather a modified effective one, as we have indicated. And second, there may be quantum corrections arising from the representation as operators of the factors that depend on the FLRW geometry in the tensor and MS Hamiltonians. In principle, all these corrections might break the invariance of the mode equations under changes of the fiducial volume. However, it is conceptually and physically most reasonable that a quantization of the system based on the variables ${\cal V}$ and $B$ should respect this invariance, since the fiducial cell must not play any physical role, but only provide an auxiliary structure to define the passage to Quantum Mechanics properly. So, as far as the quantum representation of the FLRW geometry and the definition of the peaked quantum state are independent of ${\cal V}_0$, as it is natural to require, the time-dependent masses of the perturbations would not include this fiducial volume. In this sense, the situation with respect to the invariance of the mode equations under a change of fiducial cell that we found in the classical formulation must be maintained.

It may be helpful to show, for the sake of an example, the explicit form of the time-dependent masses in a specific quantum implementation. Let us consider e.g. the so-called hybrid approach to LQC \cite{hybrinf2,hybrmass}. The mass of the tensor modes is \cite{hybrmass} \begin{equation}\label{hybridtensormass}
{\cal V}^{2/3} \left[ \frac{\sin^2{(4\pi G \gamma \sqrt{\Delta_g} B)}}{ \gamma^2 \Delta_g} - 8\pi G  V (\phi)\right] ,
\end{equation} with $\gamma$ and $\Delta_g$ defined in footnote \ref{Immi}, whereas the additional MS potential of the scalar time-dependent mass is modified to
\begin{equation}
\label{hybrMSpotential}
{ \bar{U}_{\rm MS} }={\cal V}^{2/3} \left[V_{,\phi\phi}(\phi)+ \frac{8\pi G \gamma\sqrt{\Delta_g}  \pi_{\phi}} {{\cal V}} \frac{\sin(8\pi G \gamma \sqrt{\Delta_g} B)} {\sin^2 (4\pi G \gamma \sqrt{\Delta_g} B) } V_{,\phi}(\phi)+48\pi G V(\phi)- \frac{128 \pi^2 G^2 \gamma^2 \Delta_g} { \sin^{2} (4\pi G \gamma \sqrt{\Delta_g} B)} V^2 (\phi)\right].
\end{equation}
A similar behavior with respect to the independence on ${\cal V}_0$ is also reached if one follows the dressed metric approach to LQC, even if in this latter case the quantization strategy is a little bit different, and the peak trajectory of the quantum state for the homogeneous geometry is directly lifted to the truncated perturbative phase space, without imposing a global constraint on the system that the FLRW cosmology and the perturbations form \cite{dressed2,hybrmass}.

\section{Rescaling of the physical volume and continuum limit for the modes}

We are now in an adequate position to discuss the limit of a  continuous spectrum of modes. The key observation is that the time-dependent mass of the perturbations is homogeneous of degree $2/3$ in the volume variable ${\cal V}$ in GR. This is just a consequence of the fact that the total Hamiltonian is an extensive quantity classically. If one passes to conformal time, so that the lapse becomes homogeneous of degree $1/3$ in the physical volume, the zero mode of the Hamiltonian constraint adopts correspondingly a degree equal to $2/3$. The time-dependent mass is then the wavenumber-independent part of the partial derivative of this constraint with respect to the square of the scaled configuration mode variable, which behaves essentially as an intensive variable precisely thanks to the scaling that we have chosen for it. 
	
In this classical treatment, the commented degree of homogeneity can be easily checked explicitly from the left-hand side of Eq. \eqref{tensormassderivatives} and the expression of the MS potential in Eq. \eqref{Mspotential} which, together with the homogeneous Hamiltonian \eqref{effh}, guarantee that the time-dependent mass scales with the power $2/3$ of ${\cal V}$ in GR.
	
In the effective regime of Canonical Quantum Cosmology, on the other hand, this degree of homogeneity might be altered in principle by quantum corrections. Nonetheless, it is not difficult to see that this will not happen if all functions of ${\cal V}$ are \emph{directly} represented as multiplicative operators, and the momentum $B$ acts as a derivative with respect to ${\cal V}$, perhaps up to the addition of a homogeneuos function of this volume variable of degree $-1$. Remarkably, this is the typical situation that one finds in geometrodynamics. It is probably less obvious that the same properties are maintained in other contexts for Canonical Quantum Cosmology, like the one provided by LQC. In this last case, corrections to the definition of the inverse of the volume change the degree of homogeneity in this variable \cite{ap}. Nonetheless, none of these changes affect the equations of the perturbations in the effective regime that we are studying. 

That this is the actual behavior in LQC can be easily seen for the tensor perturbations using expression \eqref{hybridtensormass}. Besides, at the order of our perturbative truncation and as far as the dynamical equations of the modes are concerned, we can use the effective constraint [obtained from Eq. \eqref{effh} by the replacement explained in footnote \ref{effect}] to check that the inflaton momentum scales like the volume ${\cal V}$. As a consequence, it follows that the MS potential \eqref{hybrMSpotential} has indeed the same behavior with respect to scalings of ${\cal V}$ as the tensor mass. On the other hand, our dynamical equations for the modes are defined with respect to a conformal time, which is obtained with a choice of the lapse function that, as we have pointed out, scales like ${\cal V}^{1/3}$. Therefore, the second derivative with respect to this time shares the homogeneity properties of the time-dependent mass, namely, it changes as ${\cal V}$ to the power $2/3$. 

Remarkably, the same result about the degree of homogeneity in ${\cal V}$ can be seen to apply to the dressed metric approach to LQC \cite{hybrmass}. In fact, this is the only important point needed for our arguments: the homogeneity in the dependence on ${\cal  V}$ of the time-dependent masses, and of the second derivative with respect to the conformal time, with homogeneity degree equal to $2/3$. Furthermore, our discussion would continue to be applicable beyond the effective regime provided that such a homogeneity is valid for the (ratios of) expectation values that determine the time-dependent masses and the change to the conformal time.

Based on our comments, if we now extract from ${\cal V}$ a (c-)number ${\cal V}_{R}$ and adjust the conformal time consequently, all the mode equations remain formally the same except for the term proportional to the square wavenumber, that gets a relative factor of ${\cal V}_{R}^{-2/3}$ compared to the rest of contributions. 
It is thus easy  to convince oneself that, by means of this rescaling of the physical volume variable, we pass in practice from wavevectors that are given by any (nonzero) integer vector $\vec{n}$ multiplied by $2\pi$ to a rescaled sequence of wavevectors of the form $2\pi \vec{n}{\cal V}_R^{-1/3}$. This rescaling can be used with a double purpose.

On the one hand, we can absorb a global scale in the definition of the physical volume. For instance, we can set the actual physical volume equal to one by choosing ${\cal V}_R$ to coincide with the (expectation) value of ${\cal V}$ at the present instant of time. Other choices are identically acceptable. To mention one example, an appealing possibility in LQC is to use this global scale to fix the expectation value of ${\cal V}$ equal to a given constant when the quantum bounce occurs in the FLRW state, namely when the minimum expectation value of ${\cal V}$ is reached.

On the other hand, we can employ the considered rescaling to reach at last the desired continuum limit for our spectrum of perturbative modes. Once we have factored out ${\cal V}_{R}$ in the volume ${\cal V}$, we can let this (c-)number tend to infinity. The spacing between components of the allowed wavevectors, equal to $2\pi {\cal V}_{R}^{-1/3}$, would then tend to zero, and the limit of $2\pi |\vec{n}| {\cal V}_{R}^{-1/3}$ for unboundedly large $|\vec{n}|$ would take all possible positive values, leading indeed to a continuous picture. From this perspective, while the description of the perturbed system has been made independent of the fiducial volume by construction, the continuum limit is attained in Fourier space by exploiting the behavior under rescalings of the physical volume. 

Let us emphasize that, in this continuum limit, the mode equations become independent of the global scale ${\cal V}_{R}$, as it should happen in the case with flat topology that we are discussing. In addition, the effective homogeneous solutions are insensitive to that scale, as far as the effective homogeneous constraint guarantees that the inflaton momentum gets rescaled in the same way as the volume ${\cal V}$. This ensures that those solutions can be constructed consistently using the constraint after having set the volume ${\cal V}$ equal to any specified constant at a given instant of time, along the lines that we explained above.

In order to check the validity of this continuum limit, let us also analyze the behavior of the tensor and MS Hamiltonians of the gauge invariant perturbations. Their respective expressions in GR are the following \cite{hybrref,hybrten}:
\begin{eqnarray}
\label{pertcons} {^{T}H_{|2}} &=& \sum_{\vec n,\epsilon}  {^{T}{H}}^{k(\vec {n}),\epsilon}_{|2} ,
\quad\quad  {^{s}H_{|2}} =\sum_{\vec n}  {^{s}H_{{|2}}^{k(\vec{n})}},  \\  \label{hybrtensham}
{^{T}{H}}^{k(\vec {n}),\epsilon}_{|2} &=& \frac{1}{2 |{\cal V}|^{1/3}}\left(\left\{k(\vec{n})^2+ 8 \pi G {\cal V}^{2/3}[2 \pi G B^2 - V(\phi)] \right\}|{\bar d}_{\vec {k},\epsilon}|^2+|\pi_{ {\bar d}_{\vec {k},\epsilon}}|^2\right), \\
\label{hybrMSham}
{^{s}H_{{|2} }^{k(\vec{n})}} &=&
\frac{1}{2 |{\cal V}|^{1/3}}\left(\left\{k(\vec{n})^2+8 \pi G {\cal V}^{2/3}[2 \pi G B^2 - V(\phi)]+ {U}_{MS} \right\}|{\bar v}_{\vec {k}}|^2+|\pi_{ {\bar v}_{\vec {k}}}|^2\right).
\end{eqnarray}
We use the notation $\vec{k}(\vec{n})=2\pi \vec{n}$, with $k(\vec{n})$ its norm, and the sum of modes is made over all (nonzero) integer vectors $\vec{n}$ and over the two polarizations in the case of the tensor modes. Accordding to our previous comments, similar expressions would be expected in Canonical Quantum Cosmology, in the effective regime that we are considering. Again for the sake of an example, in hybrid LQC the only changes are the substitution of the square of $4 \pi G \gamma \sqrt{\Delta_g} B$ by its square sinus, and of the MS potential $U_{MS}$ by its effective counterpart given in Eq. \eqref{hybrMSpotential}. If we then extract ${\cal V}_{R}$ from ${\cal V}$, as we discussed above, the wavevectors $\vec{k}(\vec{n})$ get replaced in practice by new ones, $\vec{k}_R(\vec{n})=2\pi{\vec{n}}{\cal V}_R^{-1/3}$. With this replacement, the configuration and momentum parts of the tensor and MS Hamiltonians (in the effective regime) become homogeneous functions of ${\cal V}_R$, but with different degrees of homogeneity. To get the same degree in both types of contributions, we simply redefine
\begin{equation}
\label{newmodes}
{\bar d}_{\vec k,\epsilon}= \frac{{\breve d}_{\vec k,\epsilon}}{{\cal V}_R^{1/6}},\quad\quad  {\bar v}_{\vec k}= \frac{{\breve v}_{\vec k}}{{\cal V}_R^{1/6}},\quad \quad  \pi_{{\bar d}_{\vec k,\epsilon}}= {\cal V}_R^{1/6}\pi_{{\breve d}_{\vec k,\epsilon}}, \quad \quad \pi_{{\bar v}_{\vec k}}= {\cal V}_R^{1/6} \pi_{{\breve v}_{\vec k} }.
\end{equation}
Note that this redefinition does not affect the linear equations of motion for the modes. 

In terms of these new variables for the perturbations, it is straightforward to check that the individual mode contributions to the tensor and MS Hamiltonians become homogeneous of degree equal to zero in ${\cal V}_R$. On the other hand, notice that the spacing between the components of the rescaled wavevectors $\vec{k}_R$ is $2\pi {\cal V}_R^{-1/3}$. Therefore, to transform the sum over modes in our Hamiltonians directly into a (Lebesgue) integral over real (three-dimensional) wavevectors in the continuum, the sum over $\vec{n}$ must absorb a factor $(2\pi)^{3}{\cal V}_R^{-1}$, for instance from a rescaling of the lapse function, or equivalently by redefining the time coordinate in which our Hamiltonians generate the dynamics. In this manner, one finally gets a well-defined continuum limit for the Hamiltonians of the perturbations, in which the sum over discrete modes becomes an integral over continous real modes. As a side remark, we point out that the rescaling of the lapse to produce the factor ${\cal V}_R^{-1}$ is also needed in the homogeneous sector if we want its dynamics to be independent of this (c-)number, because the effective Hamiltonian for proper time should be homogeneous of degree one in ${\cal V}$ provided that the inflaton momentum scales also as ${\cal V}$ in the analyzed trajectories (even if these did not happen to be exact homogeneous solutions).

We conclude by remarking that the restriction to compact spatial sections has fundamental advantages, while being compatible with the recovery of a well-defined continuum limit in Fourier space. This compactness ensures that the zero modes are isolated in the spectrum, allowing a distinguished treatment for them in the perturbative truncation and facilitating a way to avoid infrared divergences.

\section{Rescaling of the fiducial volume and continuum limit for the modes}

Let us now explore an alternate route to the continuum limit for the perturbative modes based on constant scale transformations of the fiducial volume. This route rests on the use of variables that do not rescale with that auxiliary volume, so that the corresponding Poisson brackets vary with changes of the fiducial structure. In fact, this alternate route has been traditional in geometrodynamics. Here, we will investigate its implementation in Canonical Quantum Cosmology and its limitations. 

If one does not use homogeneous variables that rescale with the fiducial volume, this volume can be extracted from the action as a global multiplicative factor. This factor can be absorbed by redefining the fundamental action constant of the theory (e.g. the Planck constant), so that the effective constant becomes equal to the ratio of the original one by the fiducial volume. In the kind of gravitational systems that we are studying, this has two effects. First, it produces a rescaling of the symplectic structure. Then, the variables of the system do not need to be rescaled in order to remain canonical when the fiducial volume changes, because this change is effectively absorbed in the symplectic structure, and thus in the Poisson brackets that this structure determines. Second, the Lagrange multipliers of all the constraints get redefined as well, so as to compensate the changes of the fiducial volume. 

In the case of our perturbed system, an adequate set of canonical homogeneous variables with the rescaled symplectic structure is $\{ \check{\cal V} ,B ,\phi , \check{\pi}_{\phi} \}$, where 
\begin{equation}
\label{nonchangingvariables} \check{\cal V}=\frac{{\cal V}}{{\cal V}_0},\quad\quad \check{\pi}_{\phi}=\frac{\pi_{\phi}}{{\cal V}_0}.
\end{equation}
For the gauge invariant perturbations, a careful calculation shows that the rescaling of the symplectic structure gets compensated by the passage from discrete modes, with nontrivial Poisson brackets in the form of Kronecker deltas, to continuous modes, with Dirac deltas. Keeping this result in mind, we simply change of discrete canonical pairs before taking the limit to a continuum of modes:
\begin{equation}
\label{newperturb}
\check{d}_{\vec {k},\epsilon}= {\cal V}_0^{1/6} {\bar d}_{\vec {k},\epsilon},\quad\quad \pi_{\check{d}_{\vec {k},\epsilon}}= \frac{ \pi_{\bar{d}_{\vec {k},\epsilon}} } {{\cal V}_0^{1/6}},\quad\quad \check{v}_{\vec k}= {\cal V}_0^{1/6} \bar{v}_{\vec k} ,\quad\quad  \pi_{\check{v}_{\vec k}}=\frac{ \pi_{\bar{v}_{\vec k}} } {{\cal V}_0^{1/6}}.	
\end{equation} 

Expressed in terms of these variables, the time-dependent mass is equal to ${\cal V}_0^{2/3}$ times a quantity that is independent of the fiducial volume, both for the tensor and the MS perturbations. This statement is valid, at least, for GR, for its straigthforward quantization in geometrodynamics, and for LQC in the hybrid and the dressed metric approaches. Moreover, according to the philosophy of employing quantities that are independent of the fiducial volume, let us change from the conformal time employed so far in our dynamical equations to the conformal time determined by the scale factor of the model. Notice that the difference between the old and the new conformal times is a multiplicative constant, and that the relation of the new conformal time with the proper time is independent of the fiducial volume. Our change affects the second time derivatives in our equations of harmonic-oscillator type, multiplying those derivatives by the factor ${\cal V}_0^{2/3}$, exactly as we found above for the time-dependent mass. Then, we conclude that the wavenumber that contributes with its square in our equations, once we have divided them by ${\cal V}_0^{2/3}$ to eliminate a global spurious factor, is $k=2\pi |\vec{n}|{\cal V}_0^{-1/3}$. Thus, we can  attain the continuum limit in the spectrum of the perturbations by taking the limit of infinite fiducial volume, ${\cal V}_0\rightarrow \infty$. 

Let us also consider the behavior of the different individual Hamiltonians present in the system, and let us check that, in the continuum limit, they scale in fact with the same power of the fiducial volume if we use the new variables introduced in this section. The resulting power of ${\cal V}_0$ can then be absorbed with a single, common redefinition of the lapse function. For the zero modes of the model, it is easy to see from Eq. \eqref{effh} in GR [of from its effective counterpart as long as the quantization and the state of the FLRW geometry do not introduce any dependence on the fiducial cell] that the homogeneous Hamiltonian scales like ${\cal V}_0$. In the case of the tensor and MS Hamiltonians, the expressions in GR can be found in Eqs. (\ref{pertcons})--(\ref{hybrMSham}). These expressions are independent of ${\cal V}_0$ when we use the mode variables \eqref{newperturb}, except the term that goes with the square of the wavenumber. Actually, the same must occur in Canonical Quantum Cosmology provided that the quantization procedure respects the invariance under changes of the fiducial cell, as it is reasonable to assume. We have already discussed that this is the case, for instance, in LQC in the effective regime of the hybrid and dressed metric approaches. Then, and apart from a multiplicative constant, we can reinterpret the term that depends on the wavenumber as the square norm of the wavevector $\vec{k}=2\pi \vec{n}{\cal V}_0^{-1/3}$, which varies with the fiducial volume. Nonetheless, in spite of the apparent independence on ${\cal V}_0$ of the rest of individual terms, the sums over integer vectors $\vec{n}$ need to absorb a factor ${\cal V}_0^{-1}$ to have a well-defined continuum limit, that reproduces the integrals over real modes $\vec{k}$, as we explained in the previous section. In this way, we conclude that the Hamiltonians of the gauge invariant perturbations have a well-defined homogeneity behavior with respect to ${\cal V}_0$ in the continuum limit, which is indeed the same that we obtained for the homogeneous Hamiltonian, namely a linear rescaling.

As a possible shortcoming of the use of rescalings of the fiducial volume to pass to the continuum limit in Fourier space, we note that, in addition to the conceptual tension implied by the introduction of a symplectic structure that varies effectively with the fiducial scale, one still has to deal with rescaling transformations of the physical volume in the system reached in the continuum, since these transformations have not been taken into account yet.

\section{Conclusion}

We have considered primordial perturbations around a flat FLRW cosmology and discussed the route to the continuum for the modes of those perturbations. In the framework of Canonical Quantum Cosmology, we have considered the most interesting and studied case of states for the homogeneous FLRW cosmology that are peaked on some sort of effective regime. In particular, we have commented the application of our arguments to traditional canonical quantum GR and to the specific nonperturbative program provided by LQC, both in the so-called hybrid and dressed metric approaches. In our analysis of the continuum limit for the perturbative modes, two kinds of transformations have been employed in one way or another, namely, rescalings of the fiducial volume of the cell where the spatial integrations are performed, and rescalings of its physical volume, which are independent of the former transformations because they involve changes in the scale factor (or, equivalently, of the densitized triad of the homogeneous model).

We have shown that one can indeed use variables ${\cal V}$ and $B$ so that the formulation of the homogeneous system becomes totally independent of the fiducial cell without the need of an unwanted redefinition of the symplectic structure as one changes the fiducial volume. Actually, this is true not only for the homogeneous sector of the system, but also for its gauge invariant perturbations. Nonetheless, if one starts with perturbations defined in the fiducial cell, the description that one gets leads to perturbative modes with a discrete spectrum that is in fact independent of the fiducial volume. Therefore, strictly speaking, one would not reach a continuum spectrum by letting this volume tend to infinity. Thus, one arrives to an apparent tension in the perturbative formulation, unless one defends that the treatment of the homogeneous sector and that of the perturbations can be carried out with different restrictions on the spatial sections (taking first a finite cell for the homogeneous geometry and passing afterwards to the whole of $\mathbb{R}^3$ in the study of the perturbations), as it has been argued in the dressed metric approach to LQC \cite{dressed4}. But even so, the process to reach the continuum would result cumbersome, and its generalization to other prescriptions in Canonical Quantum Cosmology would not be clear. Besides, one would still have to analyze the reformulation of the system when the rescalings of the physical volume are taken into account. With these motivations in mind, we have explored the possibility of using precisely these latter rescalings to reach the limit of a continuous spectrum of modes for the perturbations.

We have proven that, indeed, one can employ the scale transformations of the physical volume to attain the desired continuum limit. This result is based on the homogeneity properties of the time-dependent masses of the perturbations with respect to their dependence on the volume ${\cal V}$, and on the similar homogeneity of the conformal time that is used in the dynamical equations for these perturbations. For GR and for effective regimes in Canonical Quantum Cosmology under reasonable assumptions about the adopted quantization, the mass turns out to be homogeneous of degree $2/3$ and the conformal time has homogeneity degree $-1/3$. Then, extracting a reference scale from the physical volume and letting this scale tend to infinity, we have obtained the continuum limit in Fourier space. Actually, from a mathematical viewpoint, the limit would also be well defined if the homogeneity of the time-dependent masses in Canonical Quantum Cosmology were distorted with subdominant terms with respect to the used reference scale, as one expects that it could be the case beyond the effective regime, in deeper quantum regions (see e.g. \cite{hybrref}). 

In general, the procedure that arises from our analysis is that we can start the treatment of the system by integrating the spatial dependence on a compact cell with finite physical volume and, once the quantization of the model is performed and the dynamical equations for the gauge invariant perturbations are obtained, we can use the scaling transformations of the physical volume to define the continuum limit for the perturbative modes. This way to reach the continuum limit has the extra advantage of eliminating any possible redundancy in rescalings of the physical volume, showing explicitly that the resulting continuous system is invariant under changes in the choice of a reference scale ${\cal V}_R$ for this volume. This is particularly important when discussing the freedom to set initial conditions for the corresponding dynamics, since those conditions would be independent of the reference scale, as far as physical results in the continuum limit are concerned. In addition, we have checked that the Hamiltonians that generate the evolution of the gauge invariant perturbations are also well defined in this limit.

For the sake of completeness, we have explored an alternative to define the continuum of modes that has been often employed in the geometrodynamic formulation of GR. This alternative concentrates all the attention on scaling transformations of the fiducial volume, and compensates the changes in this volume by modifying the symplectic structure conveniently. This change in the symplectic structure can be thought as the result of an effective redefinition of the fundamental action constant of the system (e.g. the Planck constant). We have checked that, using this family of symplectic structures together with variables that do not depend on the fiducial volume (for instance, ${\cal V}/{\cal V}_0$ and $B$), the perturbed system admits a well-defined continuum limit by simply letting the fiducial volume grow to infinity. One may argue that the use of a family of symplectic structures, and the fact that one would have still to cope with the rescalings of the physical volume in the continuous system, have conceptual disadvantages. Nevertheless, we think that is useful to compare this route to the Fourier continuum with the previously discussed one.

Although we have restricted our analysis to flat spatial topology, there seems to be no fundamental obstruction to start with other spatial topologies. In principle, the corresponding spatial curvature would appear in the equations of the discrete modes, modifying the contributions of the square wavenumber (see e.g. Ref. \cite{mikelthesis}). The limit of a continuum of modes should be reached in a similar way as discussed here.

In summary, for cosmological perturbations around a flat FLRW model, and at least in GR and in effective regimes of Canonical Quantum Cosmology under natural conditions, we have seen that it is possible to conciliate  the use of a finite fiducial cell, on the one hand, with the construction of a formalism that is invariant under changes in the fiducial volume of this cell and, on the other hand, with the availability of a continuum limit for the modes in which the physical volume has been rescaled, for all physical purposes, by a reference volume that can be chosen freely.      

\acknowledgments

The authors are grateful to D. Mart\'{\i}n de Blas and J. Olmedo for discussions. This work was supported by Project. No. MINECO FIS2014-54800-C2-2-P and Project. No. MINECO FIS2017-86497-C2-2-P from Spain.

\end{document}